\documentclass[review,12pt]{elsarticle}
\usepackage{lineno,hyperref}
\usepackage{soul}
\modulolinenumbers[5]
\usepackage{graphicx}
\usepackage{bm}
\usepackage{amsmath,amssymb}
\usepackage{xcolor}

\usepackage[b]{esvect}
\usepackage{esdiff}
\usepackage{eqparbox}  

\usepackage[labelfont=bf,font={stretch=1},font=small]{caption}

\DeclareMathOperator*{\minimize}{min}
\usepackage{subcaption}
\usepackage{float}
\usepackage{amsfonts}
\usepackage[margin=2.5cm]{geometry}
\usepackage[T1]{fontenc}
\usepackage{cleveref}
\crefname{section}{§}{§§}
\Crefname{section}{§}{§§}
\DeclareUnicodeCharacter{3B1}{\ensuremath{\alpha}}
\usepackage{xcolor}

\usepackage{amsmath}

\makeatletter
\def\ps@pprintTitle{%
 \let\@oddhead\@empty
 \let\@evenhead\@empty
 \def\@oddfoot{}%
 \let\@evenfoot\@oddfoot}
\makeatother

\begin{document}

\begin{frontmatter}
\title{Determining large-strain metal plasticity parameters using \emph{in situ}  measurements of plastic flow past a wedge}

\author{Harshit Chawla$^{1}$, Shwetabh Yadav$^{2}$, Hrayer Aprahamian$^{1}$, Dinakar Sagapuram$^{1}$}

\address{$^{1}$Department of Industrial and Systems Engineering, Texas A\&M University, College Station, TX, USA}
\address{$^{2}$Department of Civil Engineering, Indian Institute of Technology Hyderabad, India}

\begin{abstract}
\noindent

 We present a novel approach to determine the constitutive properties of metals under large plastic strains and strain rates that otherwise are difficult to access using conventional materials testing methods. The approach exploits large-strain plastic flow past a sharp wedge, coupled with high-speed photography and image velocimetry to capture the underlying plastic flow dynamics. The inverse problem of estimating material parameters from the flow field is solved using an iterative optimization procedure that minimizes the gap between internal and external plastic work. A major advantage of the method is that it neither makes any assumptions about the flow nor requires computational  simulations. To counter the problem of non-unique parameter estimates, we propose a parameterization scheme that takes advantage of the functional form of the constitutive model and reformulates the problem into a more tractable form to identify plasticity parameters uniquely. We present studies to illustrate the principle of the method with two materials with widely different plastic flow characteristics: copper (strain hardening) and a lead-free solder alloy (rate sensitive and deformation history dependent). The results demonstrate the efficacy of the method in reliably determining the material parameters under high strain/strain rate conditions of relevance to a range of practical engineering problems.
\end{abstract}
\end{frontmatter}

\section{Introduction}
\label{sec:introduction}
There are several important technological problems where accurate modeling (e.g., using finite element analysis) of plastic deformation behavior of materials under large strain and strain rate conditions is critical for reliable design, manufacturing and performance evaluation of components. These include manufacturing processes such as machining, forming, solid-state welding and surface deformation-based processes \cite{shaw, backofen1972deformation}; defense applications (e.g., armors, penetrators, ballistic impact) \cite{salvado2017review}; and crashworthiness testing of vehicles and energy-absorbing structures. In these applications, the material deformation is well into the plastic regime and often characterized by extreme strains of several hundreds to thousand percent, strain rates in the range of $10$ to $10^6$ /s, and significant plastic heating. Other examples which involve large plastic deformation and flow, albeit over small localized volumes, include tribological contacts \cite{bowden1964friction}, erosion \cite{winter1974solid}, indentation  \cite{tabor2000hardness, chaudhri1996subsurface}, and deformation fields in the vicinity of a fast-running ductile crack or a shear band \cite{cottrell_hirsch_1976, zener1944effect, sagapuram2018evidence, viswanathan2020shear}. For computational finite element models to provide accurate predictions of the material behavior and flow kinematics, it is imperative that the parameters of the material’s constitutive law – the law that relates the flow stress to plastic strain, strain history, strain rate, and temperature – be calibrated under relevant large deformation conditions.

Conventional tension, torsion, and compression tests using cross-head devices are typically conducted at strain rates lower than $10^{-1}$ /s, with the material failure occurring at relatively low strains due to material instability (e.g., necking, barrelling); these tests are therefore not suitable for large strain/strain rate testing of materials. The most popular technique used for testing and calibration of constitutive parameters under high rates is the split Hopkinson pressure bar (SHPB) technique \cite{field2004review,salvado2017review}. While this method has the advantage of facilitating a relatively straightforward analysis of the stress-strain response due to homogeneous uniaxial stress state, and has been successfully applied to a wide range of materials, it is still limited in terms of the plastic strains (up to 0.5), strain rates (10 to $10^4$ /s) and the stress states it can achieve. As a result, material behavior at higher strains and strain rates or under complex multi-axial loading conditions is often predicted using extrapolated parameters, even though studies have shown that such extrapolation can result in poor predictions \cite{fernandez2019statistical}. Another important drawback of homogeneous deformation experiments is that determining the entire parameter set of a constitutive model is  time-consuming, especially in the case of physics-based constitutive equations with a large number of unknown parameters. This is because the calibration procedure requires multiple experiments to be carried out to provide adequate data for parameter estimation. 

In contrast to SHPB,  the Taylor impact test \cite{taylor1948use} is capable of imposing large plastic strains (effective strains up to 3) under high strain rates ($\sim 10^5$ /s). However, given the spatially non-uniform plastic flow, Taylor impact test is conventionally used for validating constitutive models vis-à-vis application as a  parameter estimation method \cite{salvado2017review}. There have been also several efforts \cite{dodd1992adiabatic,kalthoff2000modes,rittel2002shear} to develop specialized sample or test geometries, with an aim  to achieve large plastic strains;  these include hat-shaped samples, compression-shear testing, punch testing, single-edge or double-edge specimens and compact forced shear specimens. Similar to the Taylor impact test, these methods are often either used for model validation or phenomenological studies; and in limited cases where they are used to infer constitutive parameters, they require the assumption of stress and strain uniformity in the deforming region. A recent development in material testing includes the use of Rayleigh-Taylor (RT) and Richtmyer-Meshkov instabilities (RMI) to probe material behavior under extreme deformation conditions \cite{prime2019tantalum}. These methods are based on a hydrodynamic instability phenomenon and involves analysis of the wave profile and perturbations in the sample surface to estimate the strength. Although impressive for their ability to access up to 10\textsuperscript{7} /s strain rates, these platforms generally require hydrocode simulations to interpret the high rate data. Therefore, similar to the classical Taylor cylinder test, they are primarily used either to estimate an ``average'' strength (obtained by adjusting the strength parameter that enters hydrocode simulations to match the experimental wave profile data); or for validation of constitutive models. 

In this study, we present a novel approach that exploits steady-state plastic flow past a wedge to quantitatively study the large deformation behavior of metals. The most attractive feature of our method when compared to conventional methods is its ability to access a wide range of deformation parameters – effective plastic strains up to 10 and strain rates from $10^{-2}$ to $10^6$ /s. We introduce an experimental-computational framework that integrates direct full-field Particle Image Velocimetry (PIV) measurements of the plastic flow with an inverse parameter identification methodology to determine the metal plasticity parameters under large-strain conditions. Unlike some of the testing methods discussed earlier, the proposed approach is free of finite element simulations and utilizes full-field flow data to calibrate the entire constitutive parameter set in a single experiment. We demonstrate the feasibility of this method with two materials/constitutive laws and present a strategy to counter the problem of non-uniqueness 
 in inverse parameter estimation.

The remainder of this article is structured as follows: Sec.~\ref{sec:exp} provides a brief background of our experimental strategy, followed by a description of the setup and imaging techniques employed for full-field measurements. The  methodology to estimate the  constitutive parameters from  flow data is described in Sec.~\ref{sec:methodology}. Implementation of the method on two different materials, one characterized by strain-hardening behavior (OFHC copper) and other characterized by a nearly perfectly plastic but highly rate-sensitive behavior (lead-free solder alloy), is presented in Sec.~\ref{sec:result}. Implications of the study and concluding
remarks are discussed in Sec.~\ref{sec:discussion} and Sec.~\ref{sec:conclusion}, respectively.

\section{Experimental} \label{sec:exp}
\subsection{Experimental Framework}
The framework that we use for characterizing the constitutive behavior of metals is two-dimensional (plane-strain) plastic flow  past a hard asymmetric wedge-shaped tool. The important parameters associated with the process are the wedge inclination angle $\alpha$, interference depth $t_0$, and the relative velocity between the wedge and the sample $V_0$; these are shown in Fig.~\ref{fig:cutting}(a). The process is characterized by a steady/quasi-steady state plastic flow that involves shearing of the material across the deformation zone,  $OA$ in Fig.~\ref{fig:cutting}(a), as it moves past the wedge tip. From a materials testing standpoint, the main attraction of this process lies in its ability to attain large plastic strains without material rupture. For example, effective plastic strains ($\varepsilon$) in the range of unity to 10 can be imposed in the deformation zone by controlling the inclination angle $\alpha$ \cite{sagapuram2020cutting}.

The advantage in terms of controlling the strain rate also becomes apparent from a simple scaling analysis. The characteristic strain rate in the deformation zone to the first order  can be given by $\varepsilon V_0/\Delta$, with $\Delta$ being the deformation zone thickness, typically 50 to 100 $\mu$m. Thus, by varying $V_0$ in the $\mu$m/s to 10 m/s range, strain rate can be varied from $\sim 10^{-2}$ /s to 10$^{6}$ /s\footnote{These high strain rates have been also experimentally confirmed, for example, see Ref.~\cite{oxley1981mechanics}}. Not only is this range for the strain rate spanning multiple orders much wider compared to that in conventional tests, but can be achieved using a single experimental configuration without specialized sample geometries or high-rate loading platforms (e.g., compressed gas guns or explosives). The distribution of temperature in the vicinity of the wedge is governed by the balance between the heat generated due to plasticity and friction, and the heat diffused away from the zone. At low speeds $\lessapprox 0.01$ m/s, the temperature rise is typically small, whereas high homologous temperatures in excess of 0.5 may be realized at higher speeds in the 1-10 m/s range \cite{shaw}. 

At least one other advantage of the proposed method is the two-dimensional nature of the plastic flow that is well-suited for direct \emph{in situ} visualization and characterization of flow field using high-speed imaging techniques. To an Eulerian observer that is fixed related to the wedge, the deformation zone is stationary, which allows sampling of large data and material volumes over an extended period. 

Even though the idea of using a sharp wedge (cutting) tool to study the properties of metals under extreme deformation conditions has been around for a long time (dating back to Shaw \cite{shaw1950quantized} and Drucker \cite{drucker1949analysis}), prior attempts to develop it as a controlled material test have been limited. It was originally suggested that by carrying out multiple experiments at different $\alpha$ and $V_0$, and using force transducer data to calculate the corresponding shear stresses acting at the deformation zone, the relationship between the effective flow stress, strain and strain rate can be developed \cite{lira1967metal, finnie1963use}. However, a limitation of this method is that for stress and strain analysis, it is necessary to assume that the deformation is simple shear along a single plane, and that the shear stress calculated from forces is constant throughout the shear plane. While this reduction of deformation zone into a shear plane may be reasonable in the case of fully strain-hardened metals, it is certainly not applicable nor appropriate for soft ductile metals where the deformation zone is quite diffused and is strongly influenced by the initial material state \cite{sagapuram2020cutting}. Moreover, this method requires multiple experiments to construct a single stress-strain curve. The approach presented here overcomes both of these limitations. 

\subsection{\emph{In Situ} Characterization of Plastic Flow}

Experiments were conducted on a displacement-controlled testbed instrumented with a piezoelectric force sensor and high-speed imaging camera (see Fig. \ref{fig:cutting}(b)). A typical experiment involves moving a rectangular workpiece sample ($\sim$ 70 mm $\times$ 40 mm $\times$ 2 mm) against a stationary wedge at a specified $V_0$ and $t_0$ using a motorized linear actuator. The wedge was made of M2 grade high-speed steel. {Two material systems -- OFHC Cu (99.99 \% purity) and a lead-free solder alloy SAC305 with a nominal composition of 96.5\% Sn, 3\% Ag and 0.5\% Cu (by weight) -- were studied.} The forces acting on the wedge were captured along two orthogonal directions (along and perpendicular to the $V_0$ direction) using a multicomponent piezoelectric dynamometer (Kistler 9129AA dynamometer) with a natural frequency $\sim 3.5$ kHz. 

In addition to forces, synchronous \emph{in situ} observations of the plastic flow around the wedge were made using a CMOS-based high-speed optical camera (pco dimax HS4) integrated with a 10X microscopic lens. The spatial and temporal resolution of this imaging system is 0.98 $\mu$m per pixel and 100 $\mu$s, respectively, with the typical field of view being 1 mm $\times$ 1 mm in size. Imaging observations were made through a transparent glass plate that was flushed against the side-surface of the sample. The use of this glass plate restricts out-of-plane flow at the sample edge and ensures that the image is focused on a single plane, and flow observations made on the side-surface are representative of the bulk (through-thickness) behavior. To obtain quantitative information of the flow in terms of displacement,  strain rate, strain history and cumulative strain fields, the acquired images were analyzed using an image correlation algorithm called particle image velocimetry (PIV) \cite{adrian2011particle}. Material sample preparation to introduce trackable markers on the sample surface and application of PIV analysis to the present configuration were described in our earlier publications \cite{yadav2020situ,yadav2020nucleation}. In brief, the analysis involves the use of an FFT-based correlation algorithm to obtain the instantaneous displacement field by performing cross-correlation between two consecutive images. 
For the $V_0$ ranges investigated in the study (1-12 mm/s), a frame capture rate of 2 kHz was found to be adequate to yield displacement data with a small error. Uncertainty in the displacement field was quantified using an \emph{a posteriori} method based on the concept of super-resolution \cite{sciacchitano2019uncertainty}. This analysis showed that the average uncertainty in the displacement field is quite low $\lessapprox 0.8$\%.

{The plastic strain field was calculated from the displacement field using the Green strain tensor formulation for plane-strain:}
{
\begin{equation}
\begin{split}
& {d\varepsilon}_{xx} = \frac{\partial u_x}{\partial x} + \frac{1}{2}\bigg[ \bigg( \frac{\partial u_x}{\partial x} \bigg)^2 + \bigg( \frac{\partial u_y}{\partial x} \bigg)^2\bigg]; \; {d\varepsilon}_{yy} = \frac{\partial u_y}{\partial y} + \frac{1}{2}\bigg[ \bigg( \frac{\partial u_x}{\partial y} \bigg)^2 + \bigg( \frac{\partial u_y}{\partial y} \bigg)^2\bigg]; \\
    & {d\varepsilon}_{xy} = \frac{{d\gamma}_{xy}}{2} = \frac{1}{2} \bigg( \frac{\partial u_x}{\partial y} + \frac{\partial u_y}{\partial x} \bigg) + \frac{1}{2} \bigg( \frac{\partial u_x}{\partial x} \frac{\partial u_x}{\partial y} + \frac{\partial u_y}{\partial x} \frac{\partial u_y}{\partial y} \bigg )
\end{split}
\end{equation}}
\noindent

\noindent {where $u_x$ and $u_y$ are the displacements along the $x$ (horizontal) and $y$ (vertical) directions, while ${d\varepsilon}_{xx}$, ${d\varepsilon}_{yy}$ and ${d\varepsilon}_{xy}$ ($ = {d\varepsilon}_{yx})$ are the individual components of the strain tensor. Note that replacing $u$ and $v$ in the above equation with the horizontal and vertical velocity components results in the strain rate tensor. Given the large deformations involved in our problem, the total strain and plastic strain are treated as equivalent (i.e., negligible elastic strain) and computed by integrating the effective plastic strain increments ($d\bar\varepsilon$) along the material path. Incremental effective plastic strain is calculated as per von Mises strain definition:}

{
\begin{equation}
\label{eq:Estrain}
   {{d \bar{\varepsilon}}} = \frac{\sqrt{2}}{3}\sqrt{({d\varepsilon}_{xx} - {d\varepsilon}_{yy})^2 + {d\varepsilon}_{xx}^2 + {d\varepsilon}_{yy}^2  + \frac{3}{2}{d\gamma}_{xy}^2 }
\end{equation}}

{As an example, Fig.~\ref{fig:brass} shows the imaging/PIV data obtained from a typical experiment with a single-phase 70/30 brass at $\alpha = 20^\circ$ and $V_0 = 4$ mm/s. The displacement, effective strain rate ($\dot{\bar\varepsilon}=d{\bar\varepsilon}/dt$) and total effective strain ($\bar\varepsilon$) fields shown in the figure were obtained by analyzing a high-speed image sequence captured at 2 kHz. The deformation zone,  region $OA$ in Fig.~\ref{fig:brass}(c) characterized by high strain rate and sharp velocity gradient, can be seen to be highly localized with a small thickness of about 40 $\mu$m. It is this sharp velocity gradient within a small region that results in high strain rates; for instance, it may be appreciated that the characteristic strain rate is in the 50-100 /s range even though $V_0$ is only a few mm/s. Figure.~\ref{fig:brass}(d) also shows that the underlying strain field is heterogeneous, characterized by spatially non-uniform strain distribution in the removed chip and effective strain levels in excess of unity.} 

\section{Inverse Parameter Estimation Methodology}\label{sec:methodology}
{This section puts forward a general approach for estimating the constitutive  parameters based on time-resolved measurements of forces and full-field strain, strain rate and temperature information. The objective of estimating the constitutive parameters is achieved using the work equilibrium principle, which for our two-dimensional case is given by \cite{prager1968}:} 

{\begin{equation}
\label{eq:WorkEq}
    \int_{R} \left( \sigma_{xx}d\varepsilon_{xx} + \sigma_{yy}d\varepsilon_{yy} + \tau_{xy}d\gamma_{xy}\right) dA  = \int_{B} \left( T_x u_x + T_y u_y \right)dS, 
\end{equation}}

\noindent {where $\sigma_{xx}$, $\sigma_{yy}$ and $\tau_{xy}$ are stress components as a function of $x$ and $y$, and $T_x$ and $T_y$ are surface tractions at the wedge-chip interface. $dA$ represents an an area element of the deformation zone denoted by $R$, while $dS$ represents a line element of the wedge-chip contact length denoted by $B$, the integration on the left being extended over $R$ and that on the right over $B$. For a sharp wedge, it is not difficult to show that the term on the right can be rewritten as $\int_{B} ( T_x u_x - T_f u_f )dS$, where the subscript $f$ denotes the wedge-chip frictional contact, with $T_f$ and $u_f$ representing the surface traction and displacement along the contact, and the negative sign arising due to displacement being in the opposite direction to the traction. Rearranging the terms in Eq.~\ref{eq:WorkEq}, and multiplying them with the sample thickness $w$ (sample dimension normal to the viewing plane), results in:}

{\begin{equation}
\label{eq:WorkEq2}
    \underbrace{\int_{R} \left( \sigma_{xx}d\varepsilon_{xx} + \sigma_{yy}d\varepsilon_{yy} + \tau_{xy}d\gamma_{xy}\right) w dA}_{d W_{def}} + \underbrace{\int_{B} \left( T_f u_f  \right)w dS}_{d W_{friction}}  = \underbrace{\int_{B} \left( T_x u_x  \right) w dS}_{d W_{total}}, 
\end{equation}}

{Note that in the above equation, the first term on the left $dW_{def}$ represents the incremental work of deformation (work associated with material plasticity in the deformation zone), the second term $dW_{friction}$ is the incremental work expended in overcoming friction at the wedge-chip contact, whereas the term on the right $dW_{total}$ is the total incremental work spent in forcing a ductile metal past a sharp wedge. Note that  the inertial forces associated with velocity change across the deformation zone are neglected in the above equation, considering that they are several orders lower than the characteristic deformation forces over the $V_0$ range investigated in this study. For example, an estimate of the inertial forces along the $V_0$ direction can be made from $\rho w t_0 {V_0}^2$ ($\rho$ being density). This shows that they are of the order of $\sim 10^{-6}$ N, while the  typical measured force values are in the $10^2$ N range.}

{In computing the $dW_{def}$ term, the elastic contributions can be safely neglected, since in a fully developed plastic zone, the plastic strain increments are very large compared to the elastic strain increments \cite{thomsen1965,johnson1983engineering}. For example, the ratio of elastic strain increment to the plastic strain increment in the plastic zone at an effective strain of unity is of the order of $10^{-3}$. } 
 {Therefore, treating the total strain increment and plastic strain increment as identical, and using the incompressibility condition, $dW_{def}$ can be given in terms of deviatoric stresses ($\sigma^\prime$) in the principal coordinate frame as:}

{\begin{equation}
\label{eq:DefWorkEq}
    dW_{def} = \int_{R} \left( \sigma_{1}^{\prime}d\varepsilon_{1} + \sigma_{2}^{\prime}d\varepsilon_{2} \right) w dA. 
\end{equation}}

{Assuming the material follows von Mises yield criterion and L\'{e}vy-Mises plastic flow rules, and remains isotropic at all times, the above equation may be rewritten as \cite{johnson1983engineering}:}

{\begin{equation}
\label{eq:DefWorkEq2}
    dW_{def} = \int_{R} \left( \bar\sigma d\bar \varepsilon \right) w dA, 
\end{equation}}

\noindent {where $\bar\sigma = 3/2({\sigma_{1}^{\prime}}^2 + {\sigma_{2}^{\prime}}^2 + {\sigma_{3}^{\prime}}^2)^{1/2}$ is known as the effective flow stress of the material and $d\bar\varepsilon$ is the incremental effective plastic strain (see Eq.~\ref{eq:Estrain}). For the most general case of thermo-viscoplastic materials, the flow stress is given by a constitutive model of the form ${\bar{\sigma}} = f(\bar{\varepsilon}, \dot{\bar{\varepsilon}}, T, \text{deformation history}, \boldsymbol{x})$, whose parameters given by the vector $\boldsymbol{x}$ are unknown and need to be determined; symbols  $\dot{\bar{\varepsilon}}$ and $T$ represent effective plastic strain rate and temperature, respectively. 
Therefore, the incremental plastic work between two consecutive images, say $j$ and $j+1$, can be written as a function of unknown material parameters and full-field data, in a discrete summation form, as:}

{\begin{equation}
\Delta W_{def,\text{ }predicted,\text{ }j}  = \sum_{i=1}^I {\bar{\sigma}}_{i,j}d {\bar{\varepsilon}}_{i,j}A_i w =  \sum_{i=1}^I f(\bar{\varepsilon}_{i,j};\dot{\bar{\varepsilon}}_{i,j};T_{i,j}|\boldsymbol{x})d {\bar{\varepsilon}}_{i,j}A_i w,
\label{eq:predicted_work}
\end{equation}}

\noindent {where subscripts $i$ and $j$  denote the full-field data and they refer to pixel $i \in \{1,\dots,I\}$ and frame $j\in\{1,\dots,J\}$, respectively (see Fig.~\ref{fig:full_field_data}); $A_i$ is the pixel area; $w$ as before is the sample thickness.}

{The stress/plastic work calculation presented above in terms of the kinematic quantities is similar to that commonly used in the Virtual Fields Method (VFM) \cite{rossi2020approximated,jones2018parameter}. With respect to the incremental friction work $dW_{friction}$, since the gradients in the displacement along the wedge contact are small, it can be written as:
\begin{equation}
d W_{friction}  = V_f dt \int_{B}  T_f w dS = F_f V_f dt,
\label{eq:friction_work}
\end{equation}
\noindent where $V_f$ is the chip velocity parallel to the wedge face and $F_f$ is the tangential friction force parallel to the wedge face given by $F_f$ ($= F_c \sin \alpha + F_t \cos \alpha$); here,  $F_c$ and $F_t$ are the experimentally measured force components acting parallel and perpendicular to the prescribed $V_0$ direction. Similarly, the total incremental work reduces to:
\begin{equation}
d W_{total}  = V_0 dt \int_{B}  T_x w dS = F_c V_0 dt.
\label{eq:total_work}
\end{equation}
Since information about all the velocity and force terms in Eqs.~\ref{eq:friction_work} and \ref{eq:total_work} is available from the PIV and force measurements, an experimental measure of the incremental plastic work may be obtained:}
{\begin{equation}\label{eq:measured_work}
\Delta W_{def,\text{ }exp,\text{ }j} = \Delta W_{total,\text{ }j} - \Delta W_{friction,\text{ }j} = (F_c V_0  - F_f V_{f}) \Delta t,
\end{equation}
\noindent where $\Delta t$ is the inter-frame time between two consecutive image frames. The objective now is to determine the constitutive model parameter values $\boldsymbol{x}$ that, for every pair of consecutive frames ($j$ and $j+1$), leads to a predicted plastic work (Eq.~\ref{eq:predicted_work}) that matches closely with the experimentally measured plastic work (Eq.~\ref{eq:measured_work}). In order to measure the closeness of fit between predicted and measured plastic work, we use the sum of squared residuals (SSR), one of the most widely adopted metrics of discrepancy in regression analysis. The parameter estimates are then obtained by minimizing the SSR (evaluated over $J$ time instances or frames), which leads to the following optimization problem:}
{\begin{equation}
\minimize_{\boldsymbol{x}} \quad \text{SSR}=\sum_{j=2}^J\left(\sum_{i=1}^I f(\bar{\varepsilon}_{i,j};\dot{\bar{\varepsilon}}_{i,j};T_{i,j}|\boldsymbol{x})d \bar{\varepsilon}_{ij}A_i w -\Delta W_{def,\text{ }exp,\text{ }j}\right)^2
\label{eq:deterministic}
\end{equation} 
It should be noted that our parameter estimation approach, although similar in some aspects to the VFM \cite{pierron2012virtual}, does not require the user to choose an ``artificial’’ virtual velocity field.}

The optimization problem in Eq.~\eqref{eq:deterministic} is an unconstrained nonlinear (not necessarily convex) programming problem. In order to solve the above optimization problem, we utilize a family of Newton-based solution approaches that are specifically tailored to handle different constitutive laws. The choice of Newton-based algorithms is motivated by their attractive features, such as the superior (quadratic) rate of convergence and minimal input parameters when compared to other algorithms like Nelder-Mead or simulation-based algorithms \cite{NocedalSpringer}. For the classical Newton's method, we start with an initial guess for the parameters and iteratively update the parameter guesses so that the objective function is minimized. The algorithm, in general, follows the following parameter sequence:
\begin{equation}
{\boldsymbol{x}} _{k+1} = {\boldsymbol{x}}_{k} + \mu_k \boldsymbol{d}_k; \quad     \boldsymbol{d}_k = H^{-1}_k g_{k},
\label{eq:iteration}
\end{equation}
\noindent where $k$ denotes the iteration number, $\mu_k$ (= $ \underset{\mu}{\arg\min} \ \text{SSR}(\boldsymbol{x}_{k+1})$) is the step size, $\boldsymbol{d}_k$ is the direction vector, and $H_k$ and $g_k$ are the Hessian and gradient of the objective function at $\boldsymbol{x_k}$, respectively. The step size, $\mu_k$, during each iteration is optimized using the bisection line search method, a commonly used line search method to estimate the step size in Newton's method \cite{NocedalSpringer}. This search method involves taking an upper and a lower bound for the step size and iteratively halving this interval until a point is identified that minimizes the objective function along the given direction. The algorithm is terminated when the Euclidean norm of the difference between two consecutive solutions  ($||\boldsymbol{x_{k+1} - x_k}||_2$) becomes less than a specified threshold value ($\approx 10^{-5}$); or alternatively, when Euclidean norm of the gradient ($||\boldsymbol{g_k}||_2$) reaches a threshold value ($\approx 10^{-6}$).

\section{Results} \label{sec:result}
{To demonstrate the feasibility of the approach (Sec.~\ref{sec:methodology}), wedge experiments were carried out with OFHC (Oxygen-Free High Thermal Conductivity) Cu and a solder alloy SAC305 (96.5\%Sn, 3.0\%Ag, 0.5\%Cu) as model material systems. These materials were chosen for their distinctive deformation response. Cu exhibits a high strain-hardening response while SAC305 is characterized by an elastic-nearly perfect plastic type behavior with high rate sensitivity. We present results with these two materials below, followed by validation of the approach using uniaxial tension experiments.}

\subsection{OFHC Copper}
Figure \ref{fig:copper_sample} shows the effective plastic strain, strain rate, and force data obtained with copper at $V_0 = 4$ mm/s. The underlying strain field is seen to be highly heterogeneous  (Fig.~\ref{fig:copper_sample}(b)), and an analysis of the streaklines has shown that this non-uniform straining arises because of a highly unsteady (non-laminar) plastic flow that is accompanied by periodic undulations in the strain rate field. Recent research \cite{udupa2017cutting} has shown that this type of non-laminar plastic flow is a general characteristic of metals having high strain-hardening capacity (e.g., Cu, Al, Ta) and is triggered by a plastic buckling-like instability mechanism. An implication of this unsteady flow for the present work is that it enables access to a range of strain/strain rate combinations in a single experiment, which is useful for calibrating the material parameters.

Based on the literature \cite{Johnson_cook}, OFHC copper is modeled by the path-independent Johnson-Cook (JC) constitutive relation, given by: 
\begin{equation}\label{Eq:Johnson-Cook}
 \bar{\sigma} = (A + B \bar{\varepsilon}^n)\bigg[1+ C \log\Big(\frac{\dot{\bar{\varepsilon}}}{\dot{\bar{\varepsilon}}_0}\Big)\bigg]\bigg[1 + \Big(\frac{T - T_{0}}{T_m - T_{0}}\Big)^m\bigg], 
 \end{equation}
 
 \noindent {where  ${\dot{\bar\varepsilon}_0}$ is a user-defined reference strain rate, $T_m$ is melting temperature, and $T_0$ is ambient (room) temperature. Material parameters $A$, $B$, $C$, $n$ and $m$ correspond to yield strength (MPa), hardening modulus (MPa), strain-rate sensitivity, strain-hardening and thermal-softening coefficients, respectively. The reference strain rate was  chosen as  10$^{-5}$/s to ensure that ${\dot{\bar\varepsilon}}/{\dot{\bar\varepsilon}_0}$ is always  greater than 1}.  Thermal imaging measurements with various metals (including Cu) have shown that the temperature rise in the deformation zone under the low $V_0$ conditions used in the study is typically no more than 10 $^\circ$C. Given this, thermal effects can be ignored and the number of parameters reduces to four: $A$, $B$, $C$ and $n$. However, a complication is that the resultant SSR objective function is non-convex (multimodal), which poses challenges in identifying the optimal parameters using common optimization algorithms. As an example, Fig.~\ref{fig:copper_contour} shows SSR for the JC law plotted as a function of $C$ and $n$ for fixed values of $A$ (= 60 MPa) and $B$ (= 260 MPa), where the contour shows the non-convex nature of the SSR (also see inset). Although off-the-shelf local search optimization algorithms result in convergence to a stationary point, they do not guarantee a solution that is optimal in the ``global’’ sense; instead, the obtained solution will most likely correspond to a local minimum in the neighborhood of the starting point. This means that the final parameter estimates not only depend on their initial guesses (starting point) but also may not represent the true material behavior. 
 
 To circumvent this numerical issue of getting stuck at a sub-optimal solution (local minimum), the non-convex problem was reformulated into a set of convex sub-problems using an approach similar to the generalized Benders decomposition \cite{geoffrion1972generalized}. This approach exploits the structure of the objective function by temporarily fixing a few variables, referred to as complicating variables, to reduce the problem to a convex quadratic program, parameterized by the complicating variable vector. For the current objective function for the JC law, this is achieved by choosing parameters $C$ and $n$ as the complicating variables, which reduces the problem to a convex quadratic problem in $A$ and $B$. To solve for these two parameters, a grid-based optimization scheme was used where complicating variables were parameterized, and optimal $A$ and $B$ determined for each combination of $C$ and $n$ by implementing the Newton's method with optimized step size (see Sec.~\ref{sec:methodology}). {A stopping criterion,  ($||\boldsymbol{x_{k+1} - x_k}||_2$) $\leq$ $10^{-5}$, was used to terminate the algorithm.}

This involved, first, generating a coarse $C$ vs. $n$ grid, with $C$ in the range of (0.001, 0.04) with a resolution of 0.001 and $n$ in the range of (0.2, 0.8) with a resolution of 0.01, and then finding the approximate $C$ and $n$ where SSR is minimum. These parameter bounds for $C$ and $n$ were chosen based on the literature-reported ranges for common ductile metals. The least SSR for this realized grid was found to be at $C=0.005$ and $n=0.55$. The $C$ vs. $n$ grid was then subsequently refined by increasing the resolution by 10-fold in the range of $C=(0.004, 0.006)$ and $n=(0.54, 0.56)$, and the process was repeated again to further improve the solution and obtain final parameter estimates. This approach resulted in the following parameter estimates: $A = 42.13$ MPa, $B = 478.8$ MPa, $C = 0.0048$ and $n = 0.561$ for copper. This analysis has considered full-field flow data (strain and strain rate) obtained from an image sequence of 450 frames (1.5 ms inter-frame time) and  the corresponding force data acquired from an  experiment conducted at $V_0$ = 4 mm/s and $\alpha=20^{\circ}$. 

Figure~\ref{fig:copper_result}(a) shows a time-plot of the incremental work, where $\Delta W_{total}$, $\Delta W_{friction}$ and $\Delta W_{def,\text{ }exp}$ ($ = \Delta W_{total} - \Delta W_{friction}$) calculated from forces are plotted over a time period. The individual blue points in the same plot represent the predicted plastic work $\Delta W_{def, \text{ }predicted}$ calculated using the full-field deformation data and the  estimated JC model parameters. The predicted plastic work  closely matches with the measured work at all the time instances, indicating that the parameters estimated are indeed the optimal parameters. Comparison with uniaxial tension tests provides another check for the parameter estimates, e.g., see Fig.~\ref {fig:copper_result}(b). The tensile stress-strain curves obtained from copper samples pre-strained to different levels using cold rolling are shown as solid black curves, whereas the equivalent stress-strain curve constructed using the optimal JC parameters at the relevant tension-test strain rate ($2.5 \times 10^{-2}$ /s) is shown as a dashed line. The predicted curve duplicates the tension test result well, both in terms of the yield stress and strain-hardening behavior.

Robustness of the proposed approach was further verified by conducting multiple experiments with different wedge angles $\alpha$ and $V_0$ conditions. Results from these experiments are summarized in Table~\ref{tab:Copper_diffExp}. The same grid-based optimization approach was used to estimate the parameters by analyzing 500 or more frames for each experimental condition. The close match between parameters from multiple experiments demonstrates geometry (e.g., wedge inclination angle) independence of the parameter estimates and suggests that the proposed experimental/numerical approach can be used to extract reliable and quantitative mechanical property information. Equally importantly, the results also show that neither steady-state  plastic flow nor a thin uniform shear plane assumption are prerequisites for the analysis. 

\subsection{{Solder Alloy (SAC305)}}
{A similar study was carried out with the lead-free solder alloy (SAC305) characterized by a relatively low melting point ($T_m= $ 217 $^\circ$C) and high rate sensitivity. Figure~\ref{fig:woods_sample} shows the PIV deformation field for this alloy. The streakline patterns suggest that the underlying plastic flow in this case is more laminar, unlike the case for Cu.  Constitutive behavior of this alloy was modeled using Anand’s viscoplastic model \cite{brown1989internal}, an internal variable-based model that is widely used to describe the rate-dependent behavior of solder alloys.} 

{Anand model is given by:
\begin{equation}\label{eq:anand}
\bar{\sigma}= cs,
\end{equation}
\noindent
where $\bar\sigma$ is the flow stress for the steady plastic flow; $s$ is the internal variable (``deformation resistance'') with the dimensions of stress; and $c$ is a function of strain rate and temperature expressed as:
\begin{equation}\label{eq:anand2}
    c = \frac{1}{\zeta} \sinh^{-1}\Big(\frac{\dot{\bar{\varepsilon}}}{A}\exp(Q/RT)\Big)^m,
\end{equation}
\noindent
where $\dot{\bar\varepsilon}$, and $T$ have their usual meanings; $R$ is the universal gas constant while the rest are all the unknown material parameters. The evolution of the internal variable $s$ is given by:
\begin{equation}\label{eq:anand3}
            \dot{s} = \bigg[h_0 \bigg |1 - \frac{s}{s^{*}} \bigg|^a \text{sign} \bigg(1 - \frac{s}{s^{*}}\bigg)\bigg] \dot{\bar{\varepsilon}}; \quad
              s^{*} = \hat{s} \Big(\frac{\dot{\bar{\varepsilon}}}{A}\exp(Q/RT)\Big)^n,
\end{equation}
\noindent where $s^{*}$ is the saturation value of the internal variable $s$ and $\hat{s}$ is deformation resistance of $s^{*}$. The hardening or softening behavior of the material is taken into account by the \emph{sign} function in the evolution equation. The model therefore has a total of 9 parameters. However, in the absence of information about flow stress dependence on temperature, the parameters $Q$ and $A$ cannot be estimated separately. Therefore, we consider a combined term $A/\exp(Q/RT)$ which we call $A^\prime$ (1/s); a similar approach was also used previously by Brown \emph{et al.} \cite{brown1989internal}. Furthermore, since it is standard to fix the value of the stress multiplier $\zeta$ at 2  \cite{grama2015identifiability}, the number of parameters to be determined is reduced to 7.}

{From Eqs.~\ref{eq:anand}-\ref{eq:anand3}, it can be seen that the flow stress value not only depends on the  instantaneous strain rate and strain but also on the deformation history of the material. In our implementation, the internal variable value $s$ at any given material point was evaluated through path integration of all the incremental changes to this variable. This path integration is similar to how the total effective plastic strain ($\bar\varepsilon$) was computed by adding the incremental effective strains ($d\bar\varepsilon$) along the material path.}

Similar to the JC model, the applicability of common optimization algorithms for solving Eq.~\ref{eq:deterministic} for the case of Anand model is again limited by the highly non-convex nature of the underlying objective function. While grid-based optimization scheme was shown to be an effective method for solving the problem to global optimality (within resolution limits set by parameterization) for the case of JC law, a similar implementation for Anand model is not practical because of the large number of parameters (7 for Anand vs. 4 for JC). This issue was addressed using a two-step approach.

{\emph{First}, sample space of the parameter set was explored in order to identify the approximate location of the global minimum by taking hundred random initial guesses for the parameters. For this, parameter bounds for the initial guesses (see Table~\ref{tab:Anand_bounds}) were chosen based on values typically reported in the literature for solder alloys \cite{andrade2007determination}. A key observation that has emerged from this analysis is the high sensitivity of the objective function to parameter $n$. In particular, when $n > 0.05$, the algorithm resulted in parameter estimates with a relatively high SSR, with parameters also being highly dependent on the initial guess – a consequence of algorithm converging to local minima. The parameter space $n < 0.05$ on other hand was characterized by low SSR values (one order lower than for $n > 0.05$) and consistent convergence to very similar, if not exactly same, parameter estimates regardless of the initial guess. Figure~\ref{fig:SSR_woods} shows the convergence of SSR for various initial guesses, demarcated based on $n$. It is clear from the convergence plots that the resultant SSR ($\sim 3\times 10^{-4}$ to $5 \times 10^{-3}$) for $n < 0.05$ is much smaller compared to the corresponding SSR range ($\sim 1 \times 10^{-3}$ to $> 4$) for $n >  0.05$. In view of this,  as a \emph{second} step, a more rigorous search was conducted in the parameter space, $n \in (0.0001, 0.05)$, by parameterizing $n$ with a resolution of 0.01. The minimization problem for each realization of $n$ was solved by taking at least twenty random initial guesses for the remaining parameters  using the Newton's method. Parameters corresponding to the least SSR from all the initial guesses were considered to be the best estimate for a particular $n$. Neighborhood of the $n$ ($= 0.01$) having the least SSR (= $3.2 \times 10^{-4}$) was then refined again 10-fold to further improve the resolution of the solution.}

{Performance behavior of the parameters estimated using the above approach (reported in Table~\ref{tab:woods_results}) is shown in Fig.~\ref{fig:Woods_result}. Figure~\ref{fig:Woods_result}(a) shows a close match between measured plastic work and predicted plastic work computed based on the best parameter estimates from an experiment at $V_0$ = 0.2 mm/s and $\alpha$ = 20$^{\circ}$, considering 150 time instances. Figure~\ref{fig:Woods_result}(b) shows the stress-strain curves predicted using the parameter estimates (dashed lines) along with the tension test data (solid lines) at different strain rates from $10^{-4}$ to $10^{-1}$ /s.  The predicted curves are seen to capture the overall behavior well, especially the nearly perfectly plastic behavior and the high strain-rate dependence, although the elastic-plastic transition part of the stress-strain curve shows a somewhat poor agreement. As a reference, Fig.~\ref{fig:Woods_result}(c) shows the stress-strain plots, where the dashed lines now are the stress-strain curves obtained by conventional curve-fitting of Anand model to the tension test data using the calibration procedure suggested by Brown \emph{et al.} \cite{brown1989internal}. Interestingly, a similar disagreement exists between the curve fits and the data near the elastic-plastic transition even in this case. Inspection of the $R^2$ value for the stress-strain curves predicted using our inverse calibration method (Fig.~\ref{fig:Woods_result}(b)) and those obtained using conventional curve-fitting procedure (Fig.~\ref{fig:Woods_result}(c)) showed that they are 0.87 and 0.92, respectively. The lower value observed for our method may be due to the difference in the strain rates at which the stress-strain response is predicted ($10^{-4}$ to $10^{-1}$ /s) and the representative strain rates (1 to 5 /s) sampled during the wedge cutting test. Considering that the stress state in cutting (predominantly shear superimposed with hydrostatic pressure) is considerably different from the uniaxial tension test and the fact that all 7 parameters are predicted using a single text, the level of agreement is very satisfactory.}

\section{Discussion}\label{sec:discussion}

The results have shown that the proposed approach  can be a useful tool for inferring the plasticity parameters of metals and calibrating constitutive models using a limited number of experiments. 
A particularly attractive and a unique feature of the experimental cutting configuration is that it enables exploration of material behavior under very large plastic strains and strain rates of interest both from a scientific perspective and to a range of practical problems (see Sec.~\ref{sec:introduction}). Replicating such deformation conditions (especially large strains) using current high-strain rate testing methods is currently a challenge. 

Our approach, which integrates direct flow observations into the parameter estimation routine, also represents a departure from previous related studies, in that it makes no assumptions about flow kinematics or the deformation zone (e.g., stress and strain uniformity), nor it requires finite element simulations. In the latter regard, it may be noted that Finite Element Model Updating (FEMU) method \cite{martins2018comparison} has emerged as a popular inverse method for material parameter estimation over the last few decades. In its most general form, FEMU  involves iteratively adjusting the unknown parameters in order to minimize the discrepancy between the computed (FE) and experimental data. While FEMU has been shown to be a  promising technique for small deformation problems (e.g., vibration testing, structural analyses), its application to large-strain plasticity problems is still a challenge. This is because of the requirement to perform full-scale FE simulations not only at each step of the iteration (to evaluate the ``gap’’ between computation and experiment), but also in between the individual iterations to calculate the optimal (steepest) descent direction. The descent direction is determined by evaluating the derivative of the cost function with respect to each unknown parameter, and since even the simplest phenomenological viscoplastic constitutive models feature 5 or more parameters \cite{grosskreutz1975constitutive}, this procedure comes at a significant computational expense. {A few other issues with FEMU, as they particularly relate to inverse material parameter estimation  using cutting, are perhaps worth pointing out. Unlike other deformation processes, cutting is characterized by material ``bifurcation’’ or separation around the tool tip. In the conventional Lagrangian FE analysis framework, this problem is often addressed by incorporating failure criteria that enforce material separation along a predetermined parting line \cite{huang1996evaluation}. The main difficulty is that not only are simulation results sensitive to the specific failure criterion (for instance, multiple criteria have been proposed for material separation based on the critical displacement, effective plastic strain, damage parameter, etc.), but these criteria also introduce additional parameters that must be calibrated in addition to the plasticity parameters. While recent developments in hybrid Lagrangian-Eulerian (Arbitrary Lagrangian-Eulerian) techniques circumvent the need for an explicit separation criterion, the issue concerning the tool-chip contact friction still remains. In the absence of knowledge about the exact distribution of contact stresses, this contact is often modeled using a Coulomb or capped-Coulomb (i.e., friction stress limited by the local shear flow stress of the material) friction models, which may only provide a first-order approximation of the contact conditions. Also, the friction coefficient is not known \emph{a priori} but must be calibrated using experiments.} 

On the numerical side, the study has also illustrated how, even in the case of a relatively simple constitutive model like Johnson-Cook, the underlying non-convex objective function (SSR) can lead to multiple or non-unique sets of material parameters, depending on the choice of the initial parameter guesses. A concern with non-unique solutions is that though they may often predict roughly equivalent macroscopic flow stress under the strain and strain rate conditions over which they are calibrated, they do not offer performance guarantees when extrapolated to regimes outside the calibrated range \cite{jones2018parameter}. Moreover, this multiplicity of parameter sets also has implications when solving boundary value problems, since they have been shown to result in very different numerical solutions for the same boundary value problem \cite{ogden2004fitting}. Addressing this issue amounts to solving the optimization problem to global optimality. Nevertheless, most inverse parameter identification methods (e.g., FEMU, VFM) \cite{martins2018comparison, pierron2012virtual}  rely on local search algorithms that are most likely to converge to a local minimum instead of the global minimum. 

{An example of one of the most commonly adopted local search algorithms for inverse estimation schemes is the Nelder-Mead. The Nelder-Mead is a derivative-free algorithm and is known to be computationally efficient, especially for problems with a large number of unknowns. However, the relevant literature has shown that the Nelder-Mead algorithm can fail to converge to a local minimum even under favorable optimization conditions (e.g., strictly convex function with only two decision variables) \cite{mckinnon1998convergence, lagarias1998convergence}. In contrast, the Newton-based algorithm explored in this study has certain advantages. For instance, the Newton's method is guaranteed to not only  converge to the global optimal solution when the objective function is convex but also continuously improve in the best descent direction (whereas the Nelder-Mead algorithm can fail to improve from one iteration to another). A comparative study between Nelder-Mead and Newton's method to estimate Anand model parameters revealed that although the Nelder-Mead algorithm was computationally more efficient, taking an average of 15 iterations (vs. 23 iterations for Newton's method) to converge to a minimum, Newton's method converged to an objective function value that was ten times lower than Nelder-Mead's solution. 
However, due to the need for calculating derivatives, Newton-based algorithms do face certain numerical issues such as ill-conditioning or singularity, although these limitations do not appear to diminish the utility of the method inasmuch as techniques to overcome them are well developed in the literature. For example, introducing LM modification along with Newton's method can significantly reduce the negative effect of ill-conditioning, while Quasi-Newton methods (e.g., Rank 2 BFGS method) can be used to avoid the need to invert the Hessian matrix (hence overcoming singularity issues) by directly approximating the inverse of the Hessian matrix \cite{NocedalSpringer}.} 

To guarantee global optimal solution under a more general setting of a non-convex objective function, Newton-based algorithms can also be potentially tailored by taking advantage of intrinsic structural property of the problem. In this paper, we presented an approach to construct such tailored algorithms and demonstrated their application in achieving the global optimum in the ``exact’’ sense for the JC model; and heuristically for the case of the Anand model. While a unified framework that guarantees true optimal solution for parameters for any general constitutive law does not exist and is still an open research area, the numerical scheme developed here has broad applicability and work is in progress by the authors to extend the approach to other constitutive models like Zerilli-Armstrong and Mechanical Threshold Stress models. {It should be also mentioned that although model parameter estimation via curve-fitting of stress-strain data from multiple uniaxial tests is considered the norm, this procedure is still subject to non-uniqueness issues. That is, the objective function (typically SSR) that curve-fitting algorithms seek to minimize can be, and often are, non-convex, which means that local optimization algorithms that are typically used for conventional curve-fitting applications cannot guarantee global optimal solution by default \cite{andrade2007determination}. This, for example, perhaps explains the broad spectrum of parameters reported for the same material, especially in the case of complex constitutive laws like the Anand model \cite{grama2015identifiability}. The numerical scheme developed in this study to achieve optimal/epsilon-optimal solutions using a tailored constitutive law-specific approach could be also of value in this regard.}

Having calibrated the constitutive parameters over large plastic strains up to 8 and strain rates up to $10^3$ /s, a logical next step is the extension to high strain rate regimes ($\geq 10^4$ /s) by scaling the flow speed $V_0$. A main difference between relatively low speeds employed in this study and high speeds is the temperature rise in the deformation zone (due to plasticity) and at the wedge-chip interface (due to friction). A necessary step in this regard is to incorporate time-resolved  thermal field measurements in tandem with the PIV flow measurements. Similarly, the effect of inertial forces, although  safely ignored at low $V_0$, cannot be readily excluded at high speeds. {In addition, a comparison of the material parameters obtained using the wedge cutting method with  other high strain rate tests like Taylor impact and RMI would be very desirable. This comparison, for example, can involve simulating the material deformation behavior in these tests using parameter estimates obtained using our method and comparing the simulation results with experiments. In fact, such cross-comparison of data from multiple high strain rate platforms is currently a topic of much interest  \cite{prime2022broad}.} 

This study has focused on soft metals capable of undergoing large plastic deformations without material instability or failure. It would be of interest to  test the general applicability of the proposed scheme on a broader class of materials with varying ductility and strain-hardening levels. A potential complication is that certain multi-phase engineering alloys (e.g., high-strength steels, Ti and Ni alloys) are prone to shear localization, especially at high $V_0$ \cite{sagapuram2016geometric,sagapuram2018evidence,viswanathan2020shear}. While this is not a fundamental limitation, in that the proposed parameter estimation method is independent of the nature of the flow, this will, however require the use of special imaging techniques with even better spatial/temporal resolution to accurately capture the localized flow; some of our preliminary work \cite{yadav2020nucleation, yadav2020situ} touches upon full-field measurement issues in the presence of shear localization.

Besides its application as a low-cost materials testing method for characterizing the mechanical behavior of metals under very large strains and strain rates, the proposed approach has implications for other domains as well. Not only it is well-suited as a high-throughput characterization method for rapid screening of materials (e.g., additively manufactured), the small localized volumes involved in wedge cutting also make the method a practical useful tool to characterize mechanical properties of surface/sub-surface regions or as a method to determine the local constitutive behavior over small volumes (e.g., joints and welds).

\section{Conclusion}\label{sec:conclusion}
This paper has described a novel framework to determine the constitutive properties of metals under large strains using high-speed photography observations of plastic flow past a sharp wedge. An attractive feature of this experimental configuration is that it allows  access to high strains (up to 1000\%) and strain rates that are highly difficult or even impossible to achieve using conventional material testing methods. The proposed approach relies on direct \emph{in situ} observations of the plastic flow and uses work balance to obtain dynamic yield strength as a function of strain,  strain rate and deformation history. 

The study demonstrates the importance of attaining the global optimal solution in inverse parameter estimation, as failing to do so can lead to poor constitutive parameter estimates. Particular attention is therefore paid to the numerical approach used for estimating parameters from  \emph{in situ} flow data. An optimization algorithm that, when tailored, compensates for non-unique parameter estimates that arise due to the highly non-convex objective function is presented. The parameterized reformulation scheme to transform the nonlinear portion of the objective function into a more tractable form described in the paper has broad applicability and could be useful also for solving other inverse problems in engineering. 

Pure copper (strain-hardening material) and a low melting-point solder alloy (elastic-nearly perfectly plastic and rate sensitive) were used to illustrate the efficacy of this coupled high-speed photography and numerical approach to obtain material constants and characterize the flow response under large plastic strains and strain rates of up to $10^3$ /s. This is demonstrated by comparing the stress-strain curves obtained using the estimated parameters with independent uniaxial deformation tests. It is further shown that the estimated material parameters are independent of the wedge geometry. 

While thermal effects on plasticity were not considered in the study, the experimental framework can be complemented with temperature field measurements to study the temperature-dependence of flow stress in addition to strain and strain rate. Lastly, the fact that strain rates in excess of $10^5$ /s can be achieved rather easily suggests new opportunities for characterization of strain-hardening behavior of metals under high strain rates.

\section*{Funding Acknowledgement}
\noindent The work was supported by the U.S. National Science Foundation through Grant CMMI-2102030.

\setcounter{section}{0}

\pagebreak

\section*{Figures}
\numberwithin{figure}{section}
\setcounter{figure}{0}
\renewcommand\thefigure{\arabic{figure}}

\begin{figure}[h] 
    \centering
    \includegraphics[width=1\linewidth]{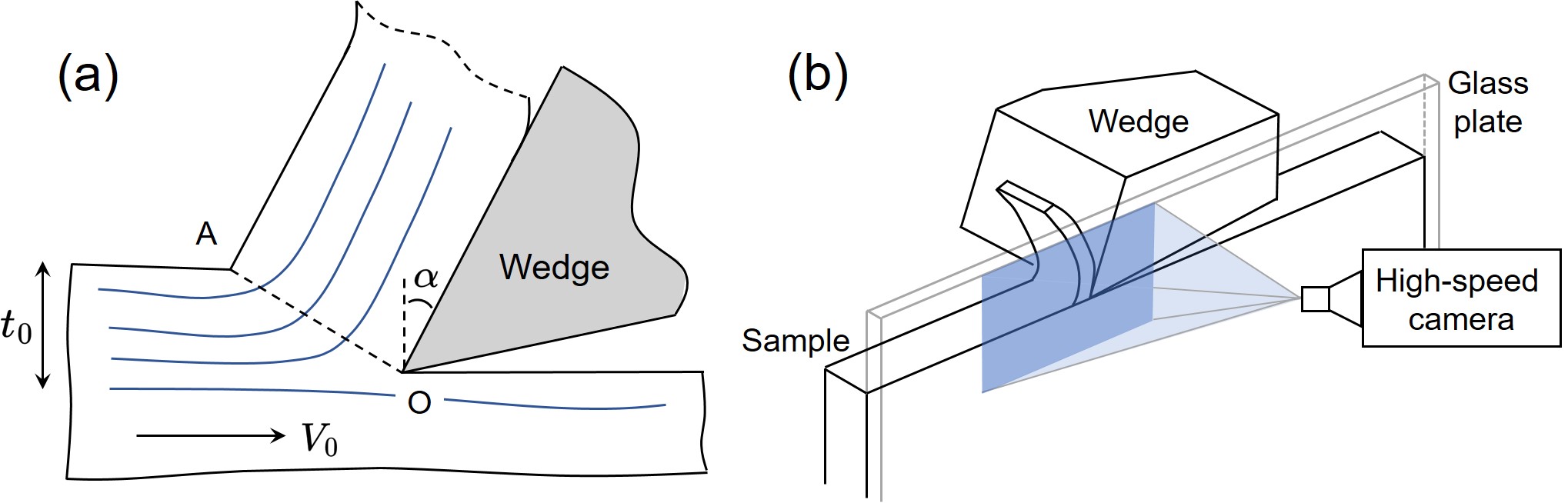}
    \caption{(a) Schematic illustrating 2D plastic flow  past a stationary wedge engaged with the sample surface. (b) Schematic of the high-speed imaging setup used for obtaining \emph{in situ} flow field data.}
    \label{fig:cutting}
\end{figure}

\begin{figure}[H]
    \centering
    \includegraphics[width=0.85\linewidth]{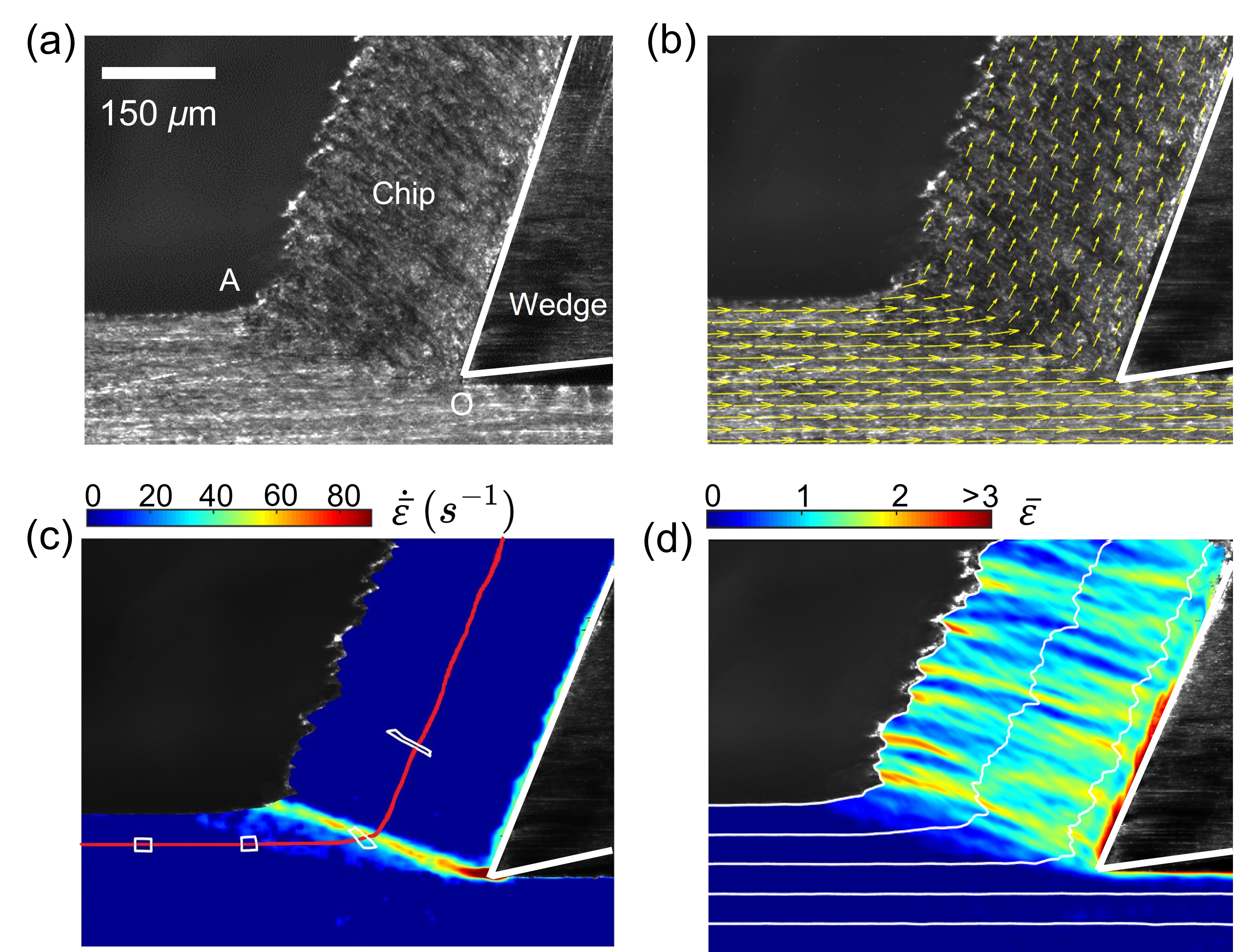}
    \caption{Sample PIV data obtained from an experiment with single-phase brass (70Cu-30Zn, wt.\%). (a) High-speed image showing  the formation of a chip. (b), (c) and (d) show the corresponding velocity,  effective (von mises) plastic strain rate $(\dot{\bar\varepsilon})$, and  effective plastic strain ($\bar \varepsilon$) fields, respectively. {(c) also shows the pathline (red line) and plastic distortion of a small material element as it passes through the deformation zone, while the white lines shown in (d) are the streaklines}. Large plastic strains in excess of unity should be noted.   $V_0 = 4$ mm/s, $\alpha = 20^\circ$ and $t_0 = 80$ $\mu$m.}
    \label{fig:brass}
\end{figure}

\begin{figure}[H]
    \centering
    \includegraphics[width=0.65\linewidth]{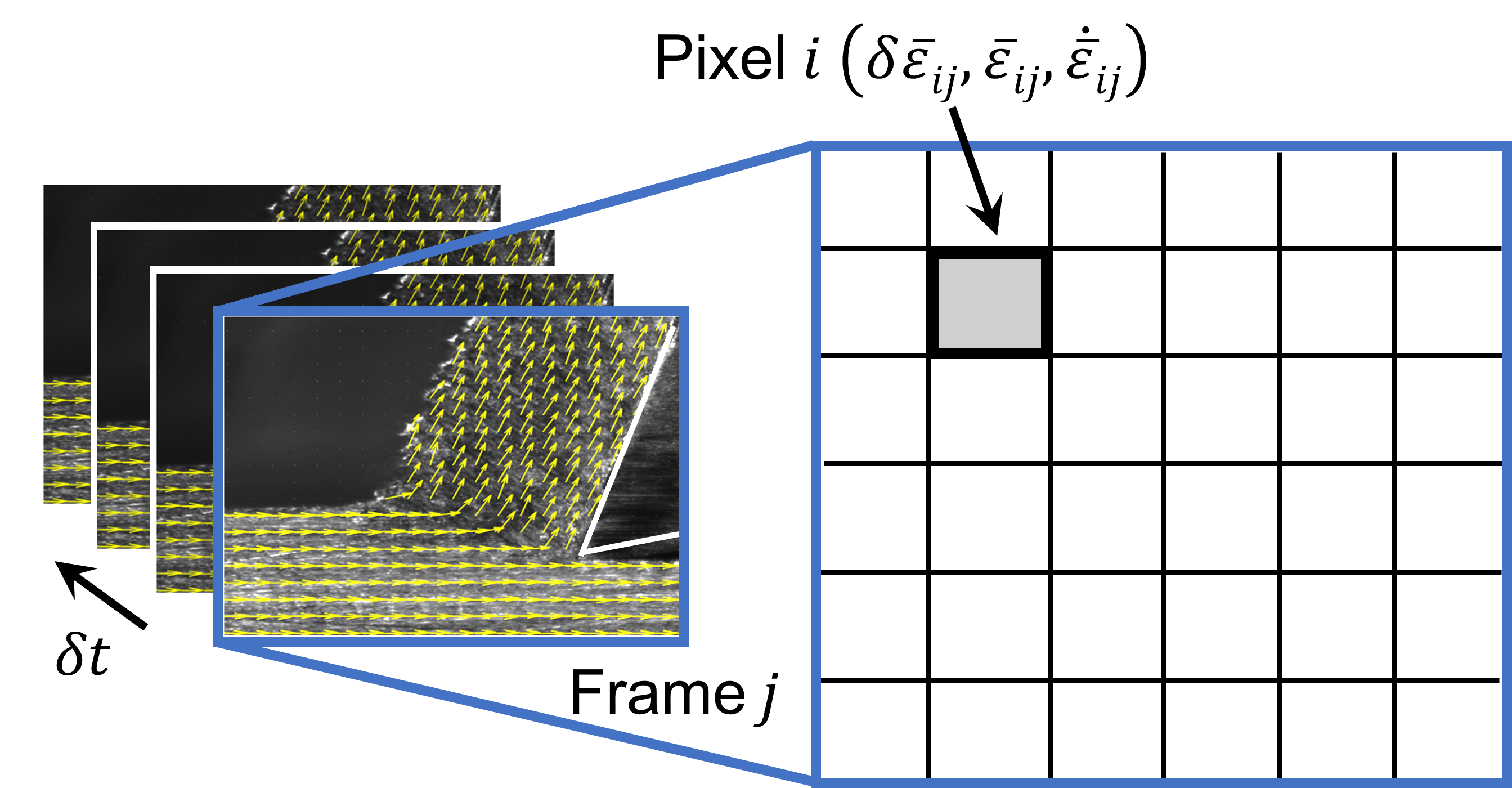}
    \caption{Flow field data notation.  $\delta \Bar{\varepsilon}_{ij}, \Bar{\varepsilon}_{ij}$, and $\dot{\Bar{\varepsilon}}_{ij}$ correspond to incremental effective plastic strain (between two consecutive images), effective plastic strain, and effective plastic strain rate at $i^{th}$ pixel location of $j^{th}$ frame or time instance.}
    \label{fig:full_field_data}
\end{figure} 

\begin{figure}[H]
    \centering
    \includegraphics[width=0.85\linewidth]{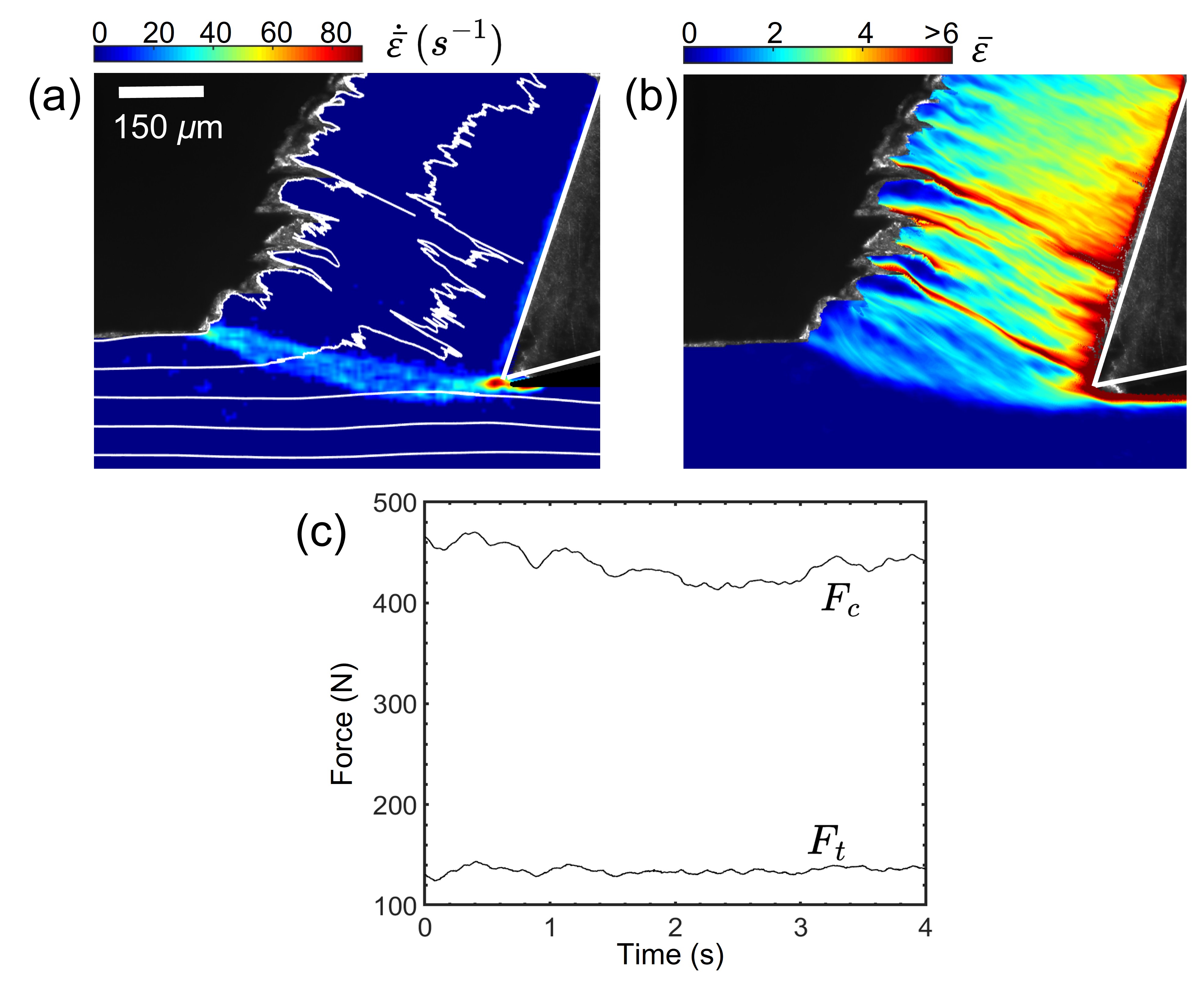}
    \caption{Plastic flow field in copper: (a) strain rate field (with superimposed streaklines) and (b) strain field showing highly unsteady (non-laminar) plastic flow and associated strain heterogeneity.  (c) Plot showing forces along two orthogonal directions:  $F_c$ (along $V_0$) and  $F_t$ (perpendicular to $V_0$). 
    $V_0 = 4$ mm/s, $\alpha = 20^\circ$ and $t_0 = 80$ $\mu$m.} 
    \label{fig:copper_sample}
\end{figure}

\begin{figure}[H]
    \centering
    \includegraphics[width=0.9\linewidth]{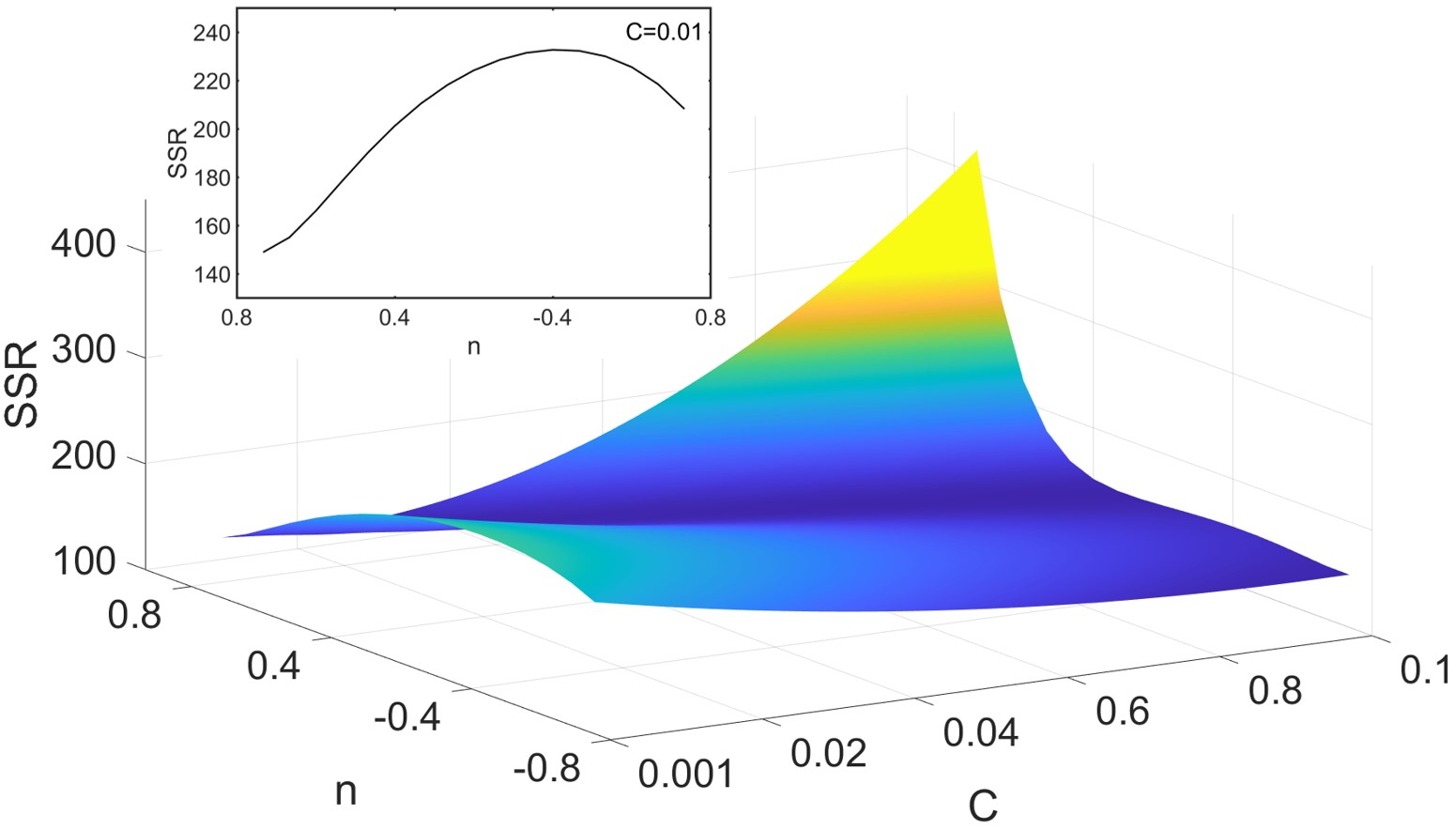}
    \caption{Objective function (SSR) contour for JC law (copper) plotted against $n$ and $C$ at $A$ = 60 MPa and $B$ = 260 MPa. The non-convex nature of the objective function is evident. Also see inset plotted at $C = 0.01$, which demonstrates solution dependence on the choice of the initial guess for the parameters.}
    \label{fig:copper_contour}
\end{figure}

\begin{figure}[H]
    \centering
    \includegraphics[width=1\linewidth]{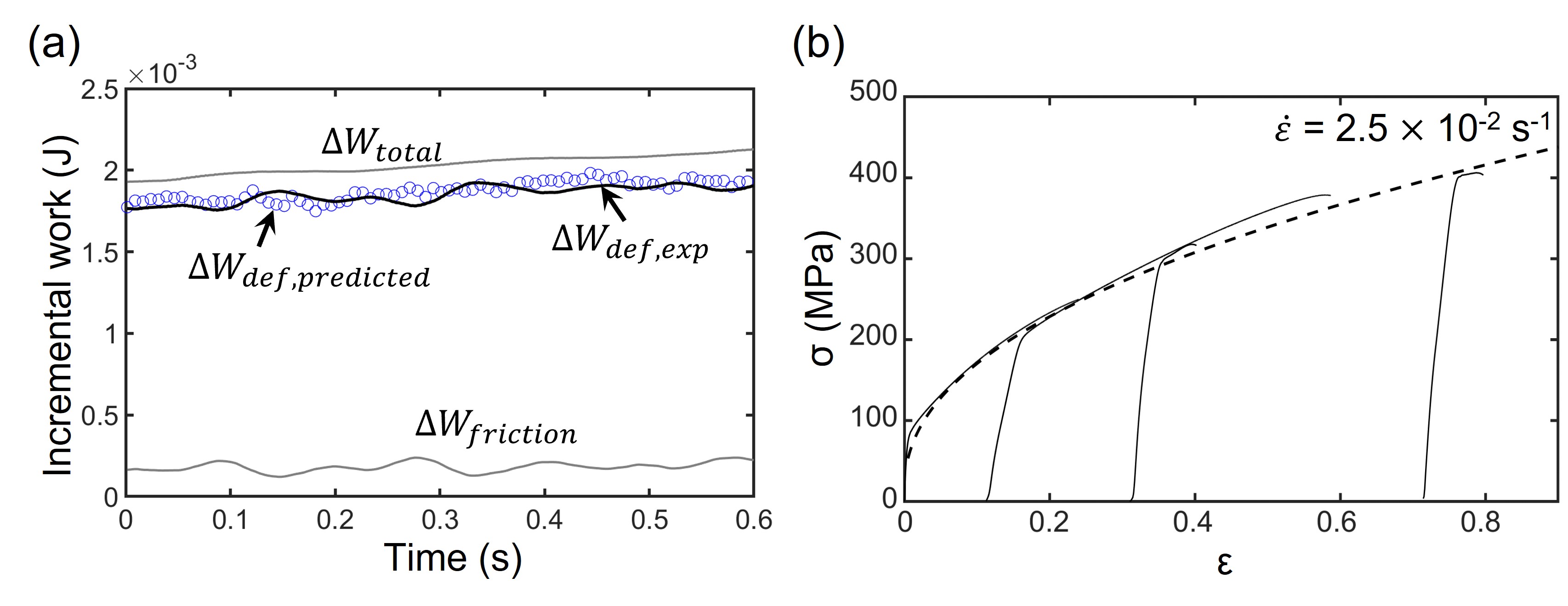}
    \caption{(a) Time-plot for copper showing  the measured plastic $\Delta W_{def,\text{ }exp}$ (black line) and the predicted plastic work $\Delta W_{def, \text{ }predicted}$ (scattered blue points) computed based on the estimated JC parameters, considering 450 frames of full-field data separated by 1.5 ms. (b) Comparison of uniaxial tensile stress-strain data (solid lines) with the corresponding curve (dashed line) obtained based on the JC parameter estimates. The different stress-strain curves (solid lines) in (b) correspond to individual tension tests  ($\dot\varepsilon = 2.5 \times 10^{-2}$ /s) performed on four copper samples with different levels of pre-strain: $\varepsilon = 0$ (annealed sample), $\varepsilon = 0.12$, $\varepsilon = 0.32$ and $\varepsilon = 0.72$. The predicted curve (dashed line) is seen to capture the strain-hardening behavior of copper very well over a wide strain range. }
    \label{fig:copper_result}
\end{figure}

\begin{figure}[H]
    \centering
    \includegraphics[width=0.85\linewidth]{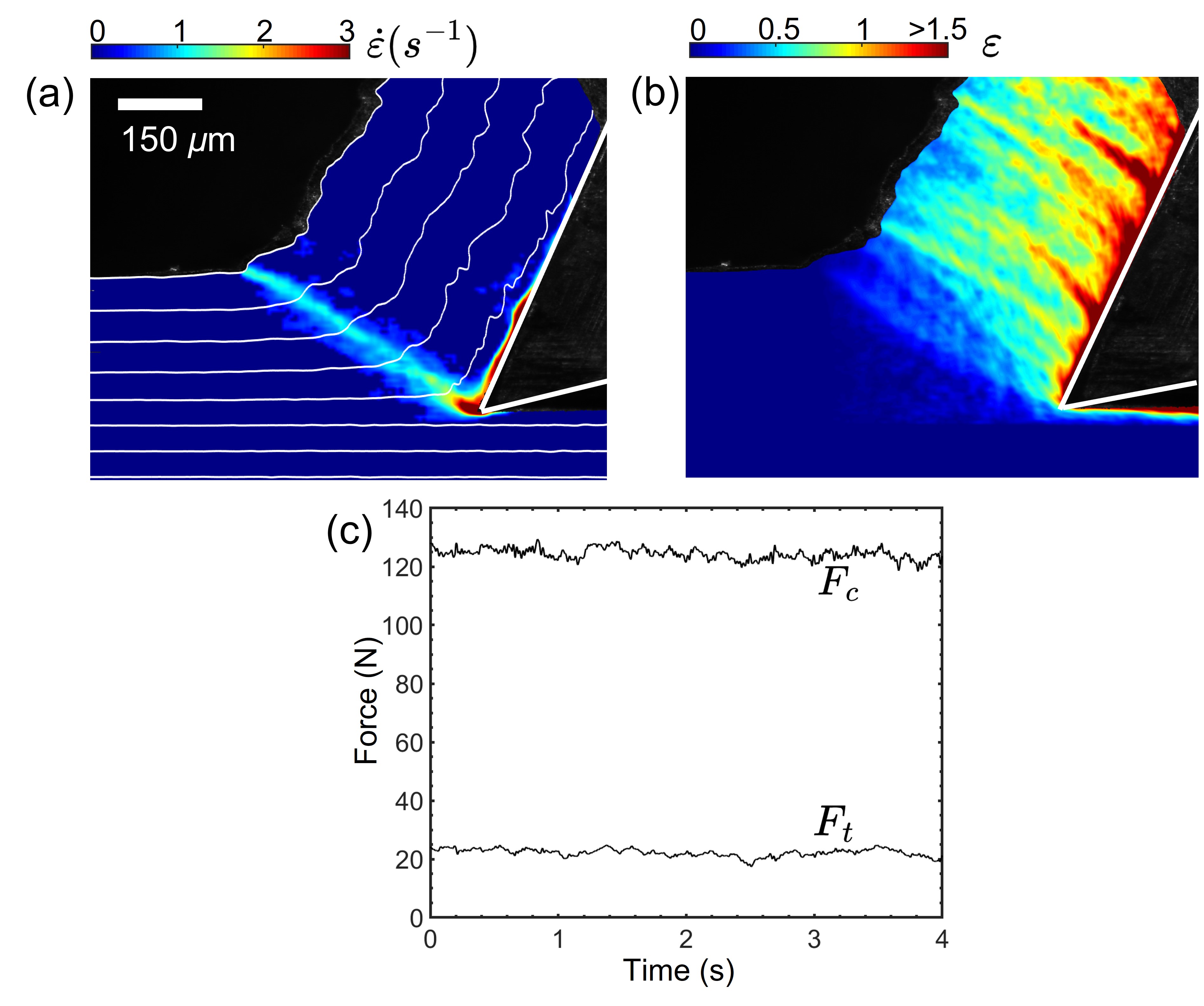}
    \caption{{Flow field in the solder alloy (SAC305): (a) effective strain rate field (with superimposed streaklines), (b)  effective strain field, and (c) forces $F_c$ and $F_t$. Note the laminar-type flow with a steady-state force profile. $V_0 = 0.2$ mm/s, $\alpha = 20^\circ$ and $t_0 = 230$ $\mu$m.}}
    \label{fig:woods_sample}
\end{figure}

\begin{figure}[H]
    \centering
    \includegraphics[width=1\linewidth]{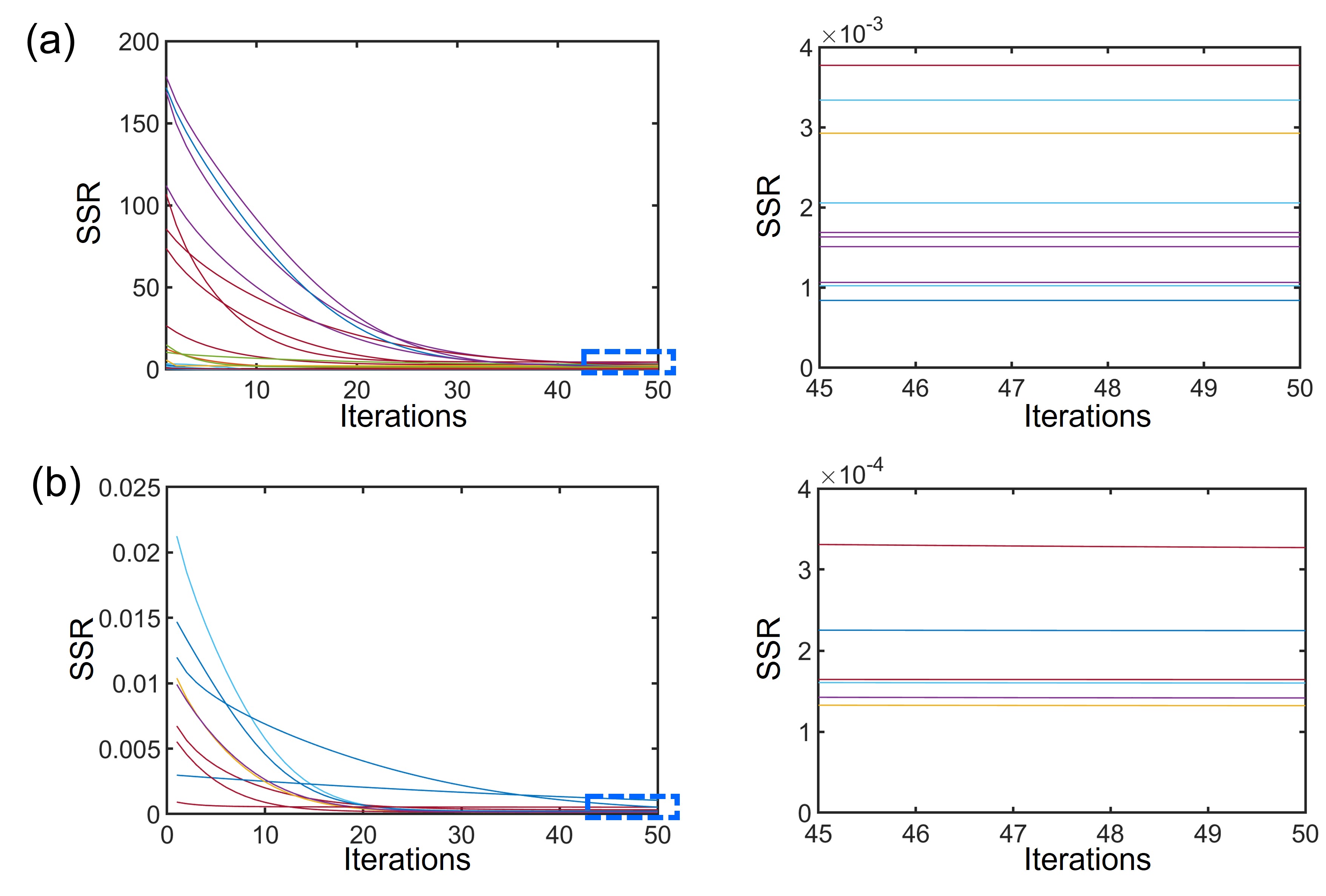}
    \caption{{Convergence behavior of the objective function (SSR) for the case of Anand law. (a) and (b) show the convergence plots (SSR vs. iterations) respectively for two cases: $n>0.05$ and $n<0.05$, where each colored line corresponds to a different initial guess chosen randomly from the parameter bounds (see Table~\ref{tab:Anand_bounds}) set for the remaining 6 parameters. The plots on right show a magnified view of the long-time behavior of the function. Objective function converges to sub-optimal solutions for all the parameter initial guesses in when $n > 0.05$, whereas optimal results with a small SSR ($\sim 10^{-4}$) are obtained for $n < 0.05$.}}
    \label{fig:SSR_woods}
\end{figure}

\begin{figure}[H] 
    \centering
    \includegraphics[width=1.0\linewidth]{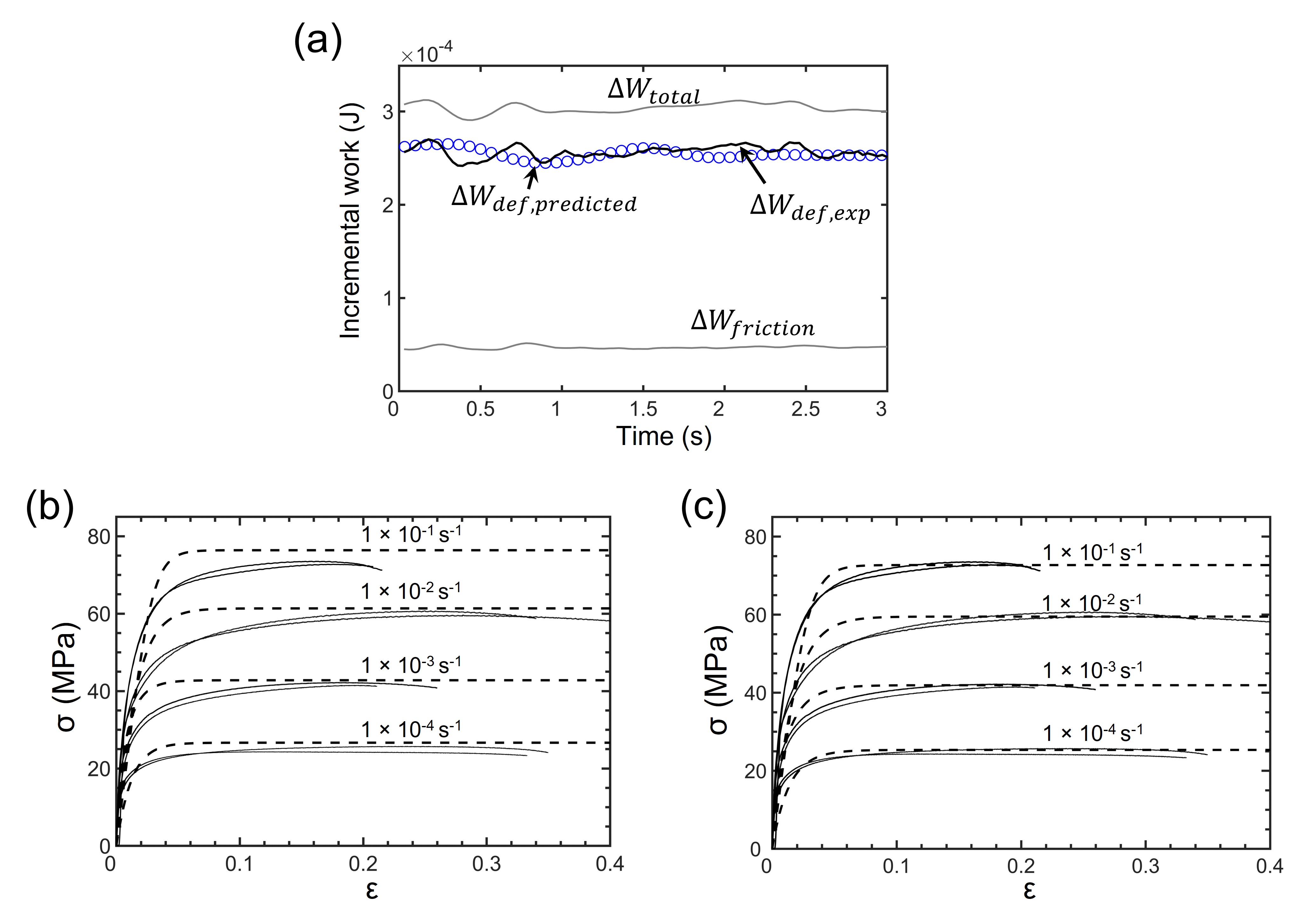}
    \caption{{(a) Time-plot for the solder alloy showing  the measured plastic work $\Delta W_{def,\text{ }exp}$ (black line) and  predicted plastic work $\Delta W_{def,\text{ }predicted}$ (blue points) calculated based on the optimal Anand parameters. A total of 150 frames (separated by 20 ms) were considered for parameter estimation.  (b) Stress-strain response of the alloy at different strain rates: solid lines are  experimental data obtained from  uniaxial tension tests, whereas dashed lines show the predicted response at equivalent strain rates based on the Anand parameters obtained using our method. (c) shows results from conventional curve-fitting of Anand model to the tension test data: solid lines are  experimental and dashed lines are the fits.}}
    \label{fig:Woods_result}
\end{figure}

\pagebreak
\section*{Tables}
\numberwithin{table}{section}
\setcounter{table}{0}
\renewcommand\thetable{\arabic{table}}
\setcounter{table}{0}

\begin{table}[h] \small
\centering
\caption{JC model parameter estimates for copper obtained from different wedge experiments.  Note the geometry-independence of material parameter estimates.}
\label{tab:Copper_diffExp}
\begin{tabular}{ccccc} 
\hline\hline
\textbf{Parameters} & \begin{tabular}[c]{@{}c@{}}~~~~$α~= 15^\circ$~~~~\\~$V_0$= 4 mm/s\end{tabular} & \begin{tabular}[c]{@{}c@{}}~~$α = 20^\circ$~~\\$V_0$ = 4 mm/s\end{tabular} & \begin{tabular}[c]{@{}c@{}}~~~~$α = 20^\circ$~~~~\\$V_0$ = 12~mm/s\end{tabular} & \begin{tabular}[c]{@{}c@{}}~~~$α = 40^\circ$~~~~\\$V_0$ = 4 mm/s\end{tabular}  \\ 
\hline
$A$ (MPa) & 40.94 & 42.13 & 45.37 & 41.61 \\ 
\hline
$B$ (MPa) & 469.2 & 478.8 & 464.8 & 471  \\ 
\hline
$C$ & 0.0052 & 0.0048 & 0.0061 & 0.0056  \\ 
\hline
$n$ & 0.549 & 0.561 & 0.525 & 0.522 \\
\hline\hline
\end{tabular}
\end{table}

\begin{table}[h] \small
\label{tab:Anand_bounds}
\centering
\caption{{Lower and upper bounds for initial guesses for Anand model parameters chosen in our parameter estimation method.}}
\label{tab:Anand_bounds}
\begin{tabular}{cccccccc} 
\hline\hline
 \textbf{Parameters} & $s_0$ & $A'$ & $m$ & $h_0$ & $s$ & $n$ & $a$ \\ 
\hline
Lower
  bound & 0 & $10^{-6}$ & 0 & 10,000 & 0 & 0 & 1 \\ 
\hline
Upper bound & 50 & $10^{-1}$ & 0.5 & 20,000 & 75 & 0.5 & 2 \\
\hline\hline
\end{tabular}
\end{table}

\begin{table}[h] \small
\centering
\caption{{Anand model parameter estimates for the lead-free solder alloy obtained from our method and conventional curve-fitting of the model to tension test data in the $10^{-4}$ to $10^{-1}$ /s strain rate range. Note that in our method all the seven parameters are determined from a single experiment.}}
\label{tab:woods_results}
\begin{tabular}{ccc}
\hline\hline
{\textbf{Parameters}} & Wedge cutting & Tension tests \\ \hline
\textbf{$s_0$} & 39.5   & 32.9  \\ \hline
\textbf{$A'$} & 7.2 $\times 10^{-4}$  & 2.8 $\times 10^{-4}$  \\ \hline
\textbf{$m$} & 0.14   & 0.113  \\ \hline
\textbf{$h_0$} & 16822 & 18793 \\ \hline
\textbf{$s$} & 59.2 & 53.8 \\ \hline
\textbf{$n$} & 0.012 & 0.011 \\ \hline
\textbf{$a$} & 1.19 & 1.04 \\ \hline\hline
\end{tabular}
\end{table}

\end{document}